\documentclass[12pt,a4paper]{article}
\pdfoutput=1
\usepackage{bm, amssymb,pifont,cancel, amsmath,comment,color}
\usepackage{here}
\usepackage{cite}
\usepackage{graphicx}
\usepackage{subfigure}

\makeatletter

\newcommand{\be}{\begin{equation}}
\newcommand{\ee}{\end{equation}}
\newcommand{\bea}{\begin{eqnarray}}
\newcommand{\eea}{\end{eqnarray}}

\def \Mpl {M_{\text{Pl}}}

\begin{document}

\begin{titlepage}

\begin{centering}
\vspace{1cm}
{\Large {\bf Continuous Spectrum on Cosmological Collider}} \\

\vspace{1.5cm}

{\bf  Shuntaro Aoki}

\vspace{.5cm}

{\it  Department of Physics, Chung-Ang University, Seoul 06974, Korea.}

\end{centering}

\vspace{2cm}

\begin{abstract}
\noindent
We study the effects of a massive field with a continuous spectrum (continuum isocurvaton) on the inflationary bispectrum in the squeezed limit. 
As a concrete example, we extend the quasi-single field inflation model to include a continuum isocurvaton with a well-motivated spectral density from extra dimensions and focus on a contribution to the bispectrum with a single continuum isocurvaton exchange. 
In contrast to the usual case without the continuous spectrum, the amplitude of the bispectrum has a damping feature in the deep squeezed limit, which can be strong evidence for the continuous spectrum.

\end{abstract}

\vspace{4cm}

\begin{flushleft} 
Email:  shuntaro@cau.ac.kr 
\end{flushleft}

\end{titlepage}

\tableofcontents
\vspace{35pt}
\hrule

\section{Introduction}\label{intro}
The cosmological collider program~\cite{Chen:2009zp,Baumann:2011nk,Noumi:2012vr,Arkani-Hamed:2015bza} is to find a new massive particle from the cosmic microwave background (CMB) observation on higher point correlation functions of the primordial curvature fluctuation (non-Gaussianity). It says that when the particle couples to inflaton, the non-Gaussianity can contain some information about the particle such as mass and spin with a specific oscillation shape (or signal) in the soft limit. In particular, the mass scale of detected particles can be as large as the Hubble scale during inflation, which is $H\sim10^{13}$ (GeV) at most. Therefore, from the cosmological observations, we can explore new physics at a very high energy scale, which cannot be reached in the terrestrial experiment. So far, a lot of works have been done in this direction~\cite{Chen:2009we,Assassi:2012zq,Sefusatti:2012ye,Norena:2012yi,Chen:2012ge,Pi:2012gf,Cespedes:2013rda,Gong:2013sma,Emami:2013lma,Kehagias:2015jha,Liu:2015tza,Dimastrogiovanni:2015pla,Schmidt:2015xka,Chen:2015lza,Delacretaz:2015edn,Bonga:2015urq,Chen:2016nrs,Flauger:2016idt,Lee:2016vti,Delacretaz:2016nhw,Meerburg:2016zdz,Chen:2016uwp,Chen:2016hrz,Kehagias:2017cym,An:2017hlx,Tong:2017iat,Iyer:2017qzw,An:2017rwo,Kumar:2017ecc,RiquelmeM:2017qhp,Franciolini:2017ktv,Tong:2018tqf,Chen:2018sce,Saito:2018omt,Cabass:2018roz,Wang:2018tbf,Chen:2018xck,Bartolo:2018hjc,Dimastrogiovanni:2018uqy,Bordin:2018pca,Chen:2018cgg,Achucarro:2018ngj,Chua:2018dqh,Kumar:2018jxz,Goon:2018fyu,Wu:2018lmx,Anninos:2019nib,Li:2019ves,McAneny:2019epy,Kim:2019wjo,Alexander:2019vtb,Lu:2019tjj,Hook:2019zxa,Hook:2019vcn,Kumar:2019ebj,Liu:2019fag,Wang:2019gbi,Wang:2019gok,Wang:2020uic,Li:2020xwr,Wang:2020ioa,Fan:2020xgh,Bodas:2020yho, Aoki:2020zbj,Maru:2021ezc,Kim:2021pbr,Lu:2021gso,Sou:2021juh, Lu:2021wxu,Wang:2021qez,Pinol:2021aun,Cui:2021iie,Tong:2022cdz,Reece:2022soh,Chen:2022vzh,Qin:2022lva,Cabass:2022rhr,Cabass:2022oap,Niu:2022quw,Niu:2022fki} with the recent development of ``cosmological bootstrap"~\cite{Arkani-Hamed:2018kmz,Sleight:2019mgd,Sleight:2019hfp,Baumann:2019oyu,Baumann:2020dch,Pajer:2020wnj,Sleight:2020obc,Goodhew:2020hob,Pajer:2020wxk,Jazayeri:2021fvk,Melville:2021lst,Goodhew:2021oqg,Sleight:2021iix,Gomez:2021qfd,Bonifacio:2021azc,Meltzer:2021zin,Hogervorst:2021uvp,DiPietro:2021sjt,Sleight:2021plv,Cabass:2021fnw,Tong:2021wai,Baumann:2021fxj,Gomez:2021ujt,Baumann:2022jpr,Heckelbacher:2022hbq,Pimentel:2022fsc,Jazayeri:2022kjy,Qin:2022fbv,Xianyu:2022jwk,Wang:2022eop,Qin:2023ejc} (see also AdS techniques~\cite{Albayrak:2018tam,Albayrak:2019asr,Albayrak:2019yve,Albayrak:2020bso,Albayrak:2020fyp}).

From viewpoints of models beyond the standard model, it is more natural that there exists more than one particle with masses around the Hubble scale, which couple to inflaton. Supergravity is a typical example, where several scalar fields acquire the so-called Hubble-induced masses by supergravity corrections~\cite{Stewart:1994ts}.
However in the cosmological collider signal, even if there are several heavy particles with couplings to inflaton, the lightest particle normally gives the most dominant contribution to the signal unless there is an unnatural hierarchy between the couplings because the Boltzmann factor suppresses the effects of the other heavy particles more strongly. Therefore, the resultant signal from several heavy particles is almost the same as that with a single particle contribution in many cases. 
Note, however, that interference effects can give non-trivial signals when the particle masses are nearly degenerate~\cite{Aoki:2020zbj, Pinol:2021aun}. 

\begin{figure}[t]
\centering
\includegraphics[width=12.0cm]{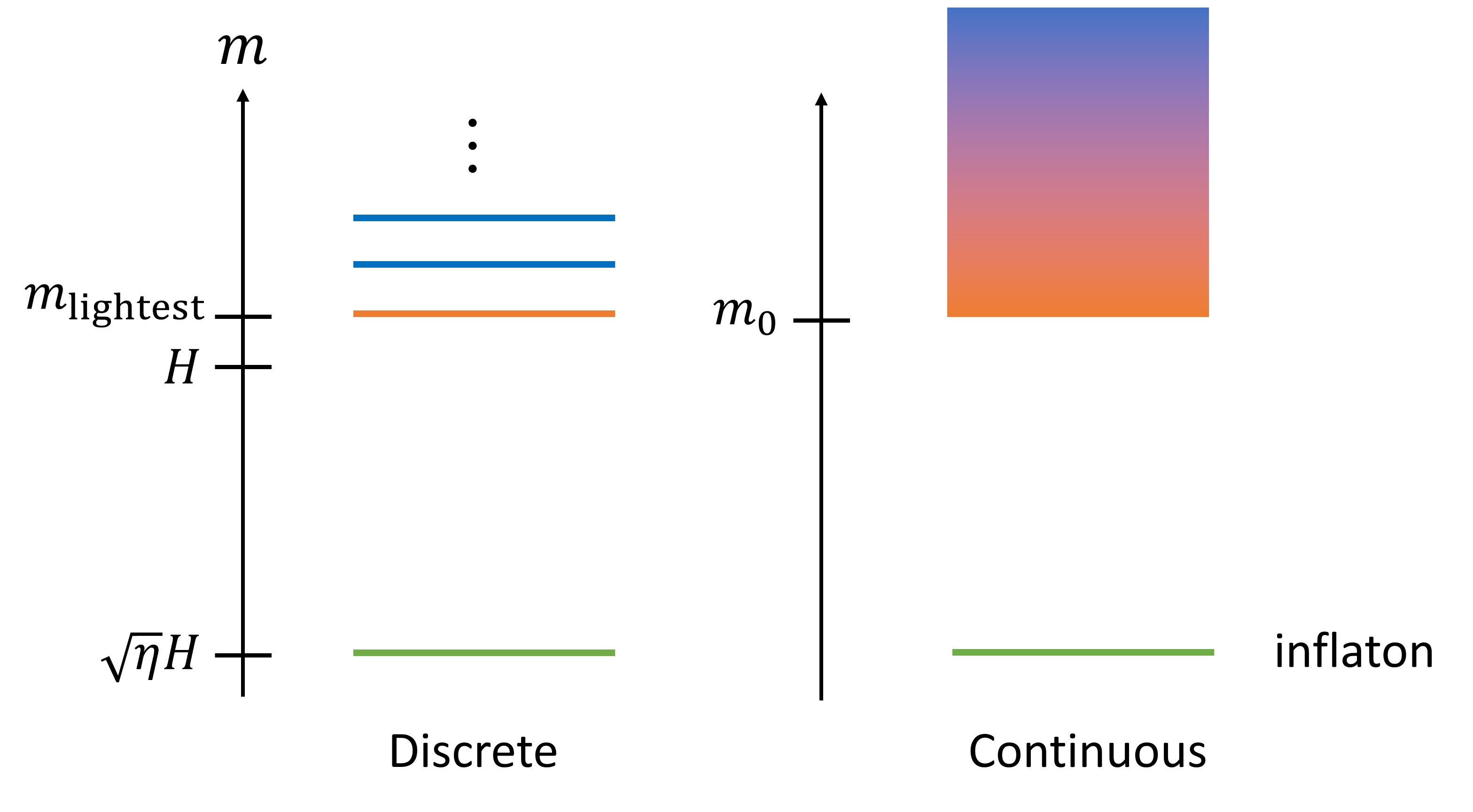}
\caption{Mass spectra for discrete and continuous cases. While the lightest particle with $m_{\rm{lightest}}$ dominates the signal in the discrete case, a certain region near the gap scale~$m_0$ collectively contributes to the signal in the continuous case.  
}
\label{fig0}
\end{figure}

In this paper, we study one of the other interesting possibilities for interference effects on the cosmological collider signal: the signal from a field with a \textit{continuous} mass spectrum. The continuous spectrum is characterized by the so-called spectral density $\rho(m^2)$ describing how the spectrum is distributed. In particular, we are interested in a ``gapped" spectral density, where $\rho(m^2)$ vanishes for $m^2<m_0^2$ with $m_0^2$ being a gap scale. It has recently been pointed out that such continuous spectra naturally emerge from warped extra dimension scenarios~\cite{Megias:2019vdb,Csaki:2021gfm,Fichet:2022ixi}, and an impact on dark matter physics was studied in Refs.~\cite{Csaki:2021gfm, Csaki:2022lnq}. Unlike the case with a discrete mass spectrum where the lightest particle determines the whole signal, in the continuous case, a certain region near the gap scale should collectively contribute to the signal, giving a non-trivial momentum dependence. The situation is schematically depicted in Fig.~\ref{fig0}. 

As a concrete setup, we adopt the quasi-single field inflation (QSFI) model~\cite{Chen:2009zp} where a massive scalar field (isocurvaton) can leave a specific signal on the three-point correlation function of the curvature perturbation (bispectrum) through kinetic couplings to inflaton. We extend the model so that it contains a massive field with a continuous spectrum instead of a fixed mass in the original model. In this paper, we call the field with a fixed mass a particle isocurvaton and that with a continuous spectrum a continuum isocurvaton, respectively. Moreover, we focus on a simple process where a single continuum isocurvaton contributes to the bispectrum, which allows us to get an analytic expression of mass-dependent coefficients of the bispectrum~\cite{Pinol:2021aun}.

The paper is organized as follows. In Sec.~\ref{setup}, we introduce the setup for the QSFI model with a continuum isocurvaton, and specify interactions between inflaton and isocurvaton fluctuations. Then, in Sec.~\ref{bispectrum}, the effects of the continuum isocurvaton on the bispectrum are investigated.  Section~\ref{Summary} is devoted to the summary. Some technicalities of the calculations are collected in Appendix~\ref{detail1} and~\ref{detail2}. 

Throughout the paper, we use the mostly plus convention $(-,+,+,+)$ for the metric. 
\section{Setup}\label{setup}
In this section, we introduce the setup. We first review the standard QSFI scenario with a particle isocurvaton~\cite{Chen:2009zp}, and then generalize it to a system with a continuum one.      

\subsection{QSFI model}\label{standard}
We consider the following action~\cite{Chen:2009zp}:
\begin{align}
S=\int d^4 x \sqrt{-g}\left[\frac{\Mpl^2}{2} R-\frac{1}{2} f(\sigma)\left(\partial_\mu \phi\right)^2-\frac{1}{2} \left(\partial_\mu \sigma\right)^2-V(\phi)-U(\sigma)\right],    \label{action}
\end{align}
where $\phi$ is an inflaton, and $\sigma$ is a particle isocurvaton which will be extended to the continuum one later. Note that $\phi$ couples to $\sigma$ through the function $f(\sigma)$.\footnote{In Ref.~\cite{Chen:2009zp}, a specific case: $f=({\rm{const.}}+\sigma)^2$ is studied, but here we keep $f$ as a general function of $\sigma$.}  Assuming the Friedman-Lemaitre-Robertson-Walker (FLRW) metric, $ds^2=-dt^2+a^2(t)d\mathbf{x}^2$, the background equations of motion are given by
\begin{align}
& 3 \Mpl^2 H^2=\frac{1}{2} f \dot{\phi}_0^2+V+U, \\
& \ddot{\phi}_0+3 H \dot{\phi}_0+\frac{1}{f} V_{\phi}=0, \\
& -\frac{f_{\sigma}}{2}\dot{\phi}_0^2+U_\sigma=0, 
\end{align}
where $H=\dot{a}/a$, and the subscript $``0"$ is attached on background fields. We also assumed that $\phi_0$ depends only on $t$ and $\sigma_0$ is a constant. The dot denotes the time derivative. The subscripts on $V,U,$ and $f$ denote the derivative with respect to the corresponding fields.

Next, we expand the fields around the background as
\begin{align}
\phi(\mathbf{x}, t)= \phi_0(t)+\delta \phi(\mathbf{x}, t), \ \     \sigma(\mathbf{x}, t)= \sigma_0+\delta \sigma(\mathbf{x}, t), 
\end{align}
with $\delta \phi$ and $\delta \sigma$ being fluctuations. Then, the free parts of Eq.~(\ref{action}) in the flat gauge are given by (see  Appendix~\ref{detail1} for details) 
\begin{align}
&\mathcal{L}_{\phi,{\rm{free}}}=\frac{a^3}{2}f(\delta \dot{\phi})^2-\frac{a}{2}  f(\partial \delta \phi)^2- a^3\left(\frac{V_{\phi\phi}}{2}-\left(3 \epsilon-\epsilon^2+\epsilon \eta\right) fH^2\right)  \delta \phi^2,\label{p_free}\\
&\mathcal{L}_{\sigma,{\rm{free}}}=\frac{a^3}{2}(\delta \dot{\sigma})^2-\frac{a}{2}  (\partial \delta \sigma)^2-\frac{a^3}{2}m^2_{\sigma}( \delta \sigma)^2,\label{s_free}
\end{align}
where $\epsilon\equiv -\frac{\dot{H}}{H^2}$ and $\eta\equiv \frac{\dot{\epsilon}}{H \epsilon} $ are slow-roll parameters, and we introduced an isocurvaton mass by $m^2_{\sigma} \equiv U_{\sigma\sigma}-\frac{1}{2}f_{\sigma\sigma}\dot{\phi}_0^2$, which is assumed to be constant in the following. The derivatives~$\partial$ appearing above are the spatial ones. Also, in the following, we ignore the mass term for $\delta \phi$, which is slow-roll suppressed, and treat it as a massless mode.

The fluctuations $\delta \phi$ and $\delta \sigma$ can be quantized by\footnote{We use a bold character such as $\boldsymbol{k}$ to denote three dimensional vectors and $k\equiv|\boldsymbol{k}|$ for their absolute
values.}
\begin{align}
& \delta \phi= \int \frac{d^3 \boldsymbol{k}}{(2 \pi)^3}\left(u_k a_{\boldsymbol{k}}+u_k^* a_{-\boldsymbol{k}}^{\dagger}\right)e^{i\boldsymbol{k} \cdot \boldsymbol{x}}, \\
& \delta \sigma=\int \frac{d^3 \boldsymbol{k}}{(2 \pi)^3}\left(v_k b_{\boldsymbol{k}}+v_k^{ *} b_{-\boldsymbol{k}}^{ \dagger} \right)e^{i\boldsymbol{k} \cdot \boldsymbol{x}},  \label{Q_sigma} 
\end{align}
where the annihilation and the creation operators, $a_{\boldsymbol{k}}, a_{\boldsymbol{k}^{\prime}}^{\dagger}, b_{\boldsymbol{k}}, b_{\boldsymbol{k}^{\prime}}^{ \dagger}$, satisfy the following commutation relations,
\begin{align}
\left[a_{\boldsymbol{k}}, a_{\boldsymbol{k}^{\prime}}^{\dagger}\right]=(2 \pi)^3 \delta^{(3)}\left(\boldsymbol{k}-\boldsymbol{k}^{\prime}\right), \quad\left[b_{\boldsymbol{k}}, b_{\boldsymbol{k}^{\prime}}^{ \dagger}\right]=(2 \pi)^3  \delta^{(3)}\left(\boldsymbol{k}-\boldsymbol{k}^{\prime}\right). \label{CR}
\end{align}
The mode functions $u_k$ and $v_k$ satisfy the equations of motions derived from Eqs.~(\ref{p_free}) and~(\ref{s_free}),
\begin{align}
& \left(a u_k\right)^{\prime \prime}+\left(k^2-\frac{2}{\tau^2}\right)\left(a u_k\right)=0, \\
& \left(a v_k\right)^{\prime \prime}+\left(k^2-\frac{2}{\tau^2}+\frac{m^2_{\sigma}}{H^2 \tau^2}\right)\left(a v_k\right)=0,
\end{align}
where $\tau$ is the conformal time, $\tau \equiv \int a^{-1} d t$, and the prime denotes the derivative with respect to $\tau$. We used a de Sitter relation, $a=-\frac{1}{H\tau}$. The solutions are given by
\begin{align}
u_k=\frac{H}{\sqrt{2f k^3}}(1+i k \tau) e^{-i k \tau}, 
\end{align}
and 
\begin{align}
v_k=-i e^{i\left(\nu_{\sigma}+\frac{1}{2}\right) \frac{\pi}{2}} \frac{\sqrt{\pi}}{2} H(-\tau)^{3 / 2} H_{\nu_{\sigma}}^{(1)}(-k \tau), \ \ {\rm{for}}\ \ m_{\sigma}<3H/2,   \label{v_nu} 
\end{align}
with $\nu_{\sigma} \equiv \sqrt{\frac{9}{4}-\left(\frac{m_{\sigma}}{H}\right)^2}$, or
\begin{align}
v_k=-i e^{-\frac{\pi}{2}\mu_{\sigma}+i\frac{\pi}{4}} \frac{\sqrt{\pi}}{2} H(-\tau)^{3 / 2} H_{i\mu_{\sigma}}^{(1)}(-k \tau), \ \ {\rm{for}}\ \ m_{\sigma}>3H/2,  \label{v_mu}  
\end{align}
with $\mu_{\sigma}=\sqrt{\left(\frac{m_{\sigma}}{H}\right)^2-\frac{9}{4}}$. Here $H_{\nu}^{(1)}(\cdot)$ is Hankel function of the first kind.

\subsection{Extension to continuum isocurvaton}
Now, let us generalize Eq.~(\ref{Q_sigma}) to a field with a continuous spectrum. This can be done by~\cite{Csaki:2021gfm},
\begin{align}
\delta \sigma=\int \frac{d m^2}{2 \pi} \sqrt{\rho\left(m^2\right)}\int \frac{d^3 \boldsymbol{k}}{(2 \pi)^3}\left(v_{k,m} b_{\boldsymbol{k},m}+v_{k,m}^{ *} b_{-\boldsymbol{k},m}^{ \dagger} \right)e^{i\boldsymbol{k} \cdot \boldsymbol{x}}, \label{dels}
\end{align}
where $\rho(m^2)$ is the spectral density, and the commutation relation~(\ref{CR}) is extended as
\begin{align}
\left[b_{\boldsymbol{k},m}, b_{\boldsymbol{k}^{\prime},m^{\prime}}^{ \dagger}\right]=(2 \pi)^4  \delta^{(3)}\left(\boldsymbol{k}-\boldsymbol{k}^{\prime}\right) \delta    \left(m^2-m^{\prime 2}\right).
\end{align}
The mode function~$v_{k,m}$ is still given by Eqs.~\eqref{v_nu} and~\eqref{v_mu}, but now characterized by a continuous mass parameter $m$ via 
$\nu =\sqrt{\frac{9}{4}-\left(\frac{m}{H}\right)^2}$ and $\mu=\sqrt{\left(\frac{m}{H}\right)^2-\frac{9}{4}}$.

Based on Eq.~\eqref{dels}, the propagator is extended to
\begin{align}
\left\langle 0\left|{\rm{T}}\left\{\delta \sigma\left(x_1\right) \delta\sigma\left(x_2\right)\right\}\right| 0\right\rangle= \int \frac{d m^2}{2 \pi} \rho\left(m^2\right) G_{++}\left(x_1, x_2; m^2\right),
\end{align}
where ${\rm{T}}(\cdots)$ is a time-ordering operator, and 
\begin{align}
\nonumber &G_{++}\left(x_1, x_2; m^2\right)\\
&\equiv \int  \frac{d^3 \boldsymbol{k}}{(2 \pi)^3}  \left(\theta\left(\tau_1-\tau_2\right)v_{k, m}\left(\tau_1\right) v_{k, m}^*\left(\tau_2\right) +\theta\left(\tau_2-\tau_1\right)v_{k, m}\left(\tau_2\right) v_{k, m}^*\left(\tau_1\right)\right)e^{i \boldsymbol{k} \cdot\left(\boldsymbol{x}_1-\boldsymbol{x}_2\right)},     
\end{align}
is the standard Schwinger-Keldysh propagator (see Ref.~\cite{Chen:2017ryl} for details). 
Here $\theta(\cdot)$ is a unit step function.
Note that the case with a  particle isocurvaton corresponds to $\rho\left(m^2\right)=2 \pi \delta\left(m^2-m_\sigma^2\right)$. 

\subsection{Interactions}
\begin{figure}[t]
\centering
\includegraphics[width=8.0cm]{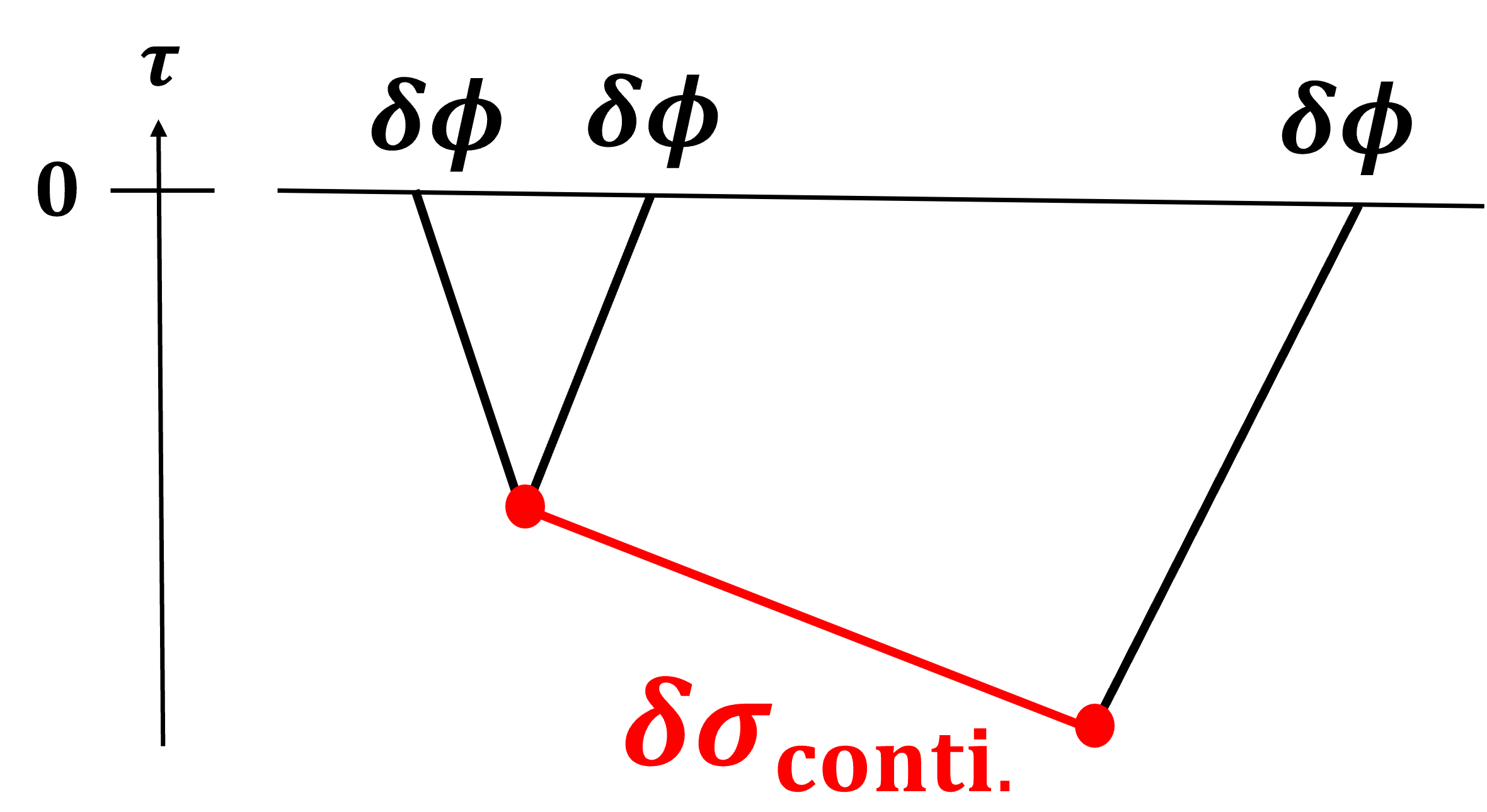}
\caption{Contribution of the continuum isocurvaton~$\delta\sigma$ to the bispectrum}
\label{fig1}
\end{figure}

Let us specify the interactions. The expanded action contains a lot of interactions between $\delta \phi$ and $\delta \sigma$. Among them, in this paper, we focus on the following interactions, 
\begin{align}
&\mathcal{L}_{2,{\rm{int}}}= a^3 f_{\sigma} \dot{\phi}_0 \delta \sigma \delta \dot{\phi},\label{L_2}\\
&\mathcal{L}_{3,{\rm{int}}}= \frac{a^3}{2}f_{\sigma}\delta \sigma \left((\delta \dot{\phi})^2-\frac{1}{a^2}(\partial \delta \phi)^2\right), \label{L_3}
\end{align}
for the continuum isocurvaton. All the quadratic and cubic interactions are shown in Appendix~\ref{detail1}.
The above terms can contribute to the bispectrum through a diagram shown in Fig.~\ref{fig1}.

We stress that the cubic interaction~(\ref{L_3}) does not necessarily give the most dominant contribution to the bispectrum. Rather, as pointed out in Ref.~\cite{Chen:2009zp}, if a cubic self interaction $\delta\sigma^3$ exists with a sizable coupling $U_{\sigma\sigma\sigma}$, it can contribute to the bispectrum with a triple exchange of $\delta \sigma$ through Eq.~(\ref{L_2}), which gives a dominant contribution to the bispectrum. Nevertheless, here we adopt Eq.~(\ref{L_3}), because one can obtain a fully analytical expression of the bispectrum in the squeezed limit (up to the integration $\int dm^2$), which greatly simplifies the analysis, but sufficient to see effects of the continuum isocurvaton.

Finally, by using a relation to the curvature perturbation $\zeta \sim -\frac{H}{\dot{\phi}_0}\delta \phi$, the interactions (\ref{L_2}) and (\ref{L_3}) can be rewritten as
\begin{align}
&\mathcal{L}_{2,{\rm{int}}}=-2a^3\epsilon\frac{f_{\sigma}}{f}\Mpl^2 H\delta \sigma\dot{\zeta},\label{L_2z}\\
&\mathcal{L}_{3,{\rm{int}}}   = a^3\epsilon \frac{f_{\sigma}}{f}\Mpl^2\delta \sigma \left(\dot{\zeta}^2-\frac{1}{a^2}(\partial \zeta)^2\right).\label{L_3z}
\end{align}

\section{Bispectrum with continuum isocurvaton}\label{bispectrum}
In this section, we study effects of the continuum isocurvaton on the bispectrum taking the process in Fig.~\ref{fig1} as an example. 

To this end, we choose a specific form of the spectral density,
\begin{align}
\rho\left(m^2\right)=\frac{2\pi}{m_0^2} \sqrt{\frac{m^2}{m_0^2}-1},  \label{SD}
\end{align}
which is studied in Refs.~\cite{Csaki:2021gfm,Csaki:2022lnq} in the context of dark matter phenomenology.
Here $m_0$ is the gap scale, from which the continuous spectrum begins. This choice of the spectral density is motivated by some warped extra dimension scenarios~\cite{Megias:2019vdb,Csaki:2021gfm}. Also, since we are interested in the oscillating signal of the bispectrum, we focus on a case with $m>m_0>3H/2$ in the following. An expression of the bispectrum with general $\rho(m^2)$ can be found in Appendix~\ref{detail2}.

The shape function $S(k_1,k_2,k_3)$ of our interest is defined through the bispectrum of $\zeta$ by
\begin{align}
\langle\zeta_{\bf{k}_1} \zeta_{\bf{k}_2}\zeta_{\bf{k}_3}\rangle =(2\pi)^7P_{\zeta}^2\frac{1}{(k_1k_2k_3)^2}\delta^{(3)}({\bf{k}}_1+{\bf{k}}_2+{\bf{k}}_3)S(k_1,k_2,k_3),    
\end{align}
where $P_{\zeta}=\frac{H^2}{8 \pi^2 \epsilon \Mpl^2}$ is the power spectrum. Here we only show the result of $S$ in the squeezed limit ($k_3\ll k_1\sim k_2$), leaving the details for Appendix~\ref{detail2}. The result is  
\begin{align}
S=-\frac{\pi}{4} \Mpl^2 \epsilon\left(\frac{f_{\sigma}}{f}\right)^2 \kappa^{-1/2}\int_{m_0^2}^{\infty} \frac{d m^2}{2 \pi} \rho\left(m^2\right)e^{-\pi\mu}{\rm{Im}}\left\{\left(\frac{1}{\kappa}\right)^{-i \mu}J_{-}(i\mu)+\left(\frac{1}{\kappa}\right)^{i \mu}J_{+}(i\mu)\right\}, \label{S}
\end{align}
where $\kappa\equiv \frac{k_1}{k_3}$, and
\begin{align}
&J_+(i\mu)= -\frac{\sqrt{\pi} 2^{-1-2 i \mu}\left(e^{\pi \mu}-i\right) \Gamma\left(i \mu+\frac{7}{2}\right)(\operatorname{csch}(\pi \mu)+\operatorname{sech}(\pi \mu))}{(2 \mu-i) \Gamma(i \mu+1)},  \\
&J_-(i\mu)=J_+(-i\mu)e^{2\pi \mu}.
\end{align}
Note that this is a natural generalization to the continuous spectrum version of the results obtained for the discrete spectrum~\cite{Pinol:2021aun}.

Defining $S=-\frac{\pi}{4} \Mpl^2 \epsilon\left(\frac{f_{\sigma}}{f}\right)^2 \kappa^{-1/2} F_{\rm{conti.}}$, we extract a purely oscillating part of the shape function with the continuum isocurvaton:
\begin{align}
 F_{\rm{conti.}} =\int_{m_0^2}^{\infty} \frac{d m^2}{2 \pi} \rho\left(m^2\right) e^{-\pi\mu} \mathcal{A}(\mu)\sin \left(\mu\log \kappa+\varphi (\mu)\right) ,\label{F}
\end{align}
where  
\begin{align}
&\mathcal{A}(\mu)=\left[\left\{{\rm{Im}}J_-(i\mu)+{\rm{Im}}J_+(i\mu)\right\}^2+\left\{{\rm{Re}}J_-(i\mu)-{\rm{Re}}J_+(i\mu)\right\}^2\right]^{1/2} ,\\
&\varphi(\mu)={\rm{arctan}}\left(\frac{{\rm{Im}}J_-(i\mu)+{\rm{Im}}J_+(i\mu)}{{\rm{Re}}J_-(i\mu)-{\rm{Re}}J_+(i\mu)}\right).
\end{align}

For comparison, here we show the result of the standard case with the particle isocurvaton,   
\begin{align}
 F_{\rm{particle}} =e^{-\pi\mu_{\sigma}} \mathcal{A}(\mu_{\sigma})\sin \left(\mu_{\sigma}\log \kappa+\varphi (\mu_{\sigma})\right) ,\label{F_normal}
\end{align}
where $\mu_{\sigma}=\sqrt{\left(\frac{m_{\sigma}}{H}\right)^2-\frac{9}{4}}$. 

\begin{figure}[t]
 \begin{minipage}{0.5\hsize}
  \begin{center}
   \includegraphics[width=70mm]{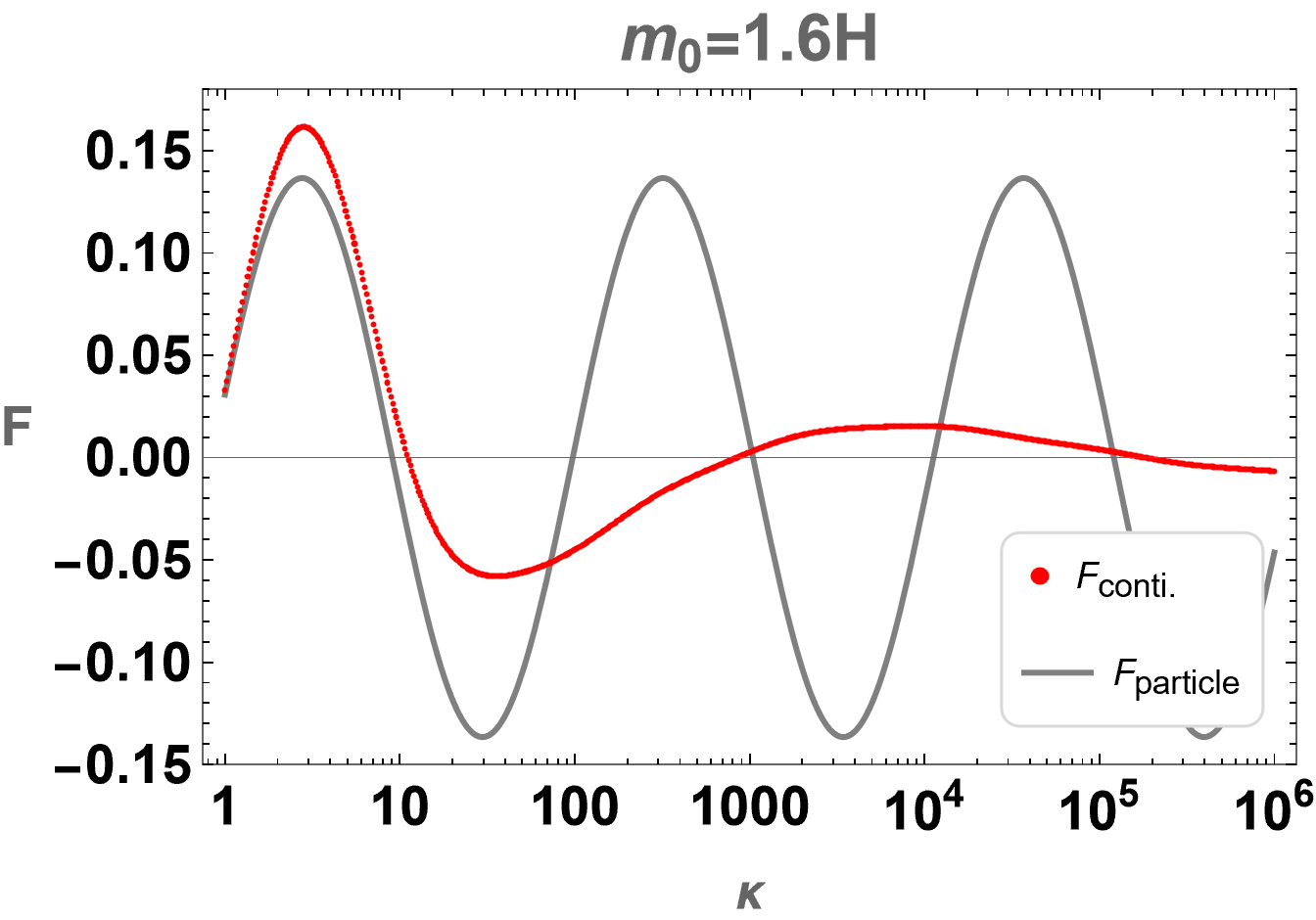}
  \end{center}
 \end{minipage}
 \begin{minipage}{0.5\hsize}
  \begin{center}
   \includegraphics[width=70mm]{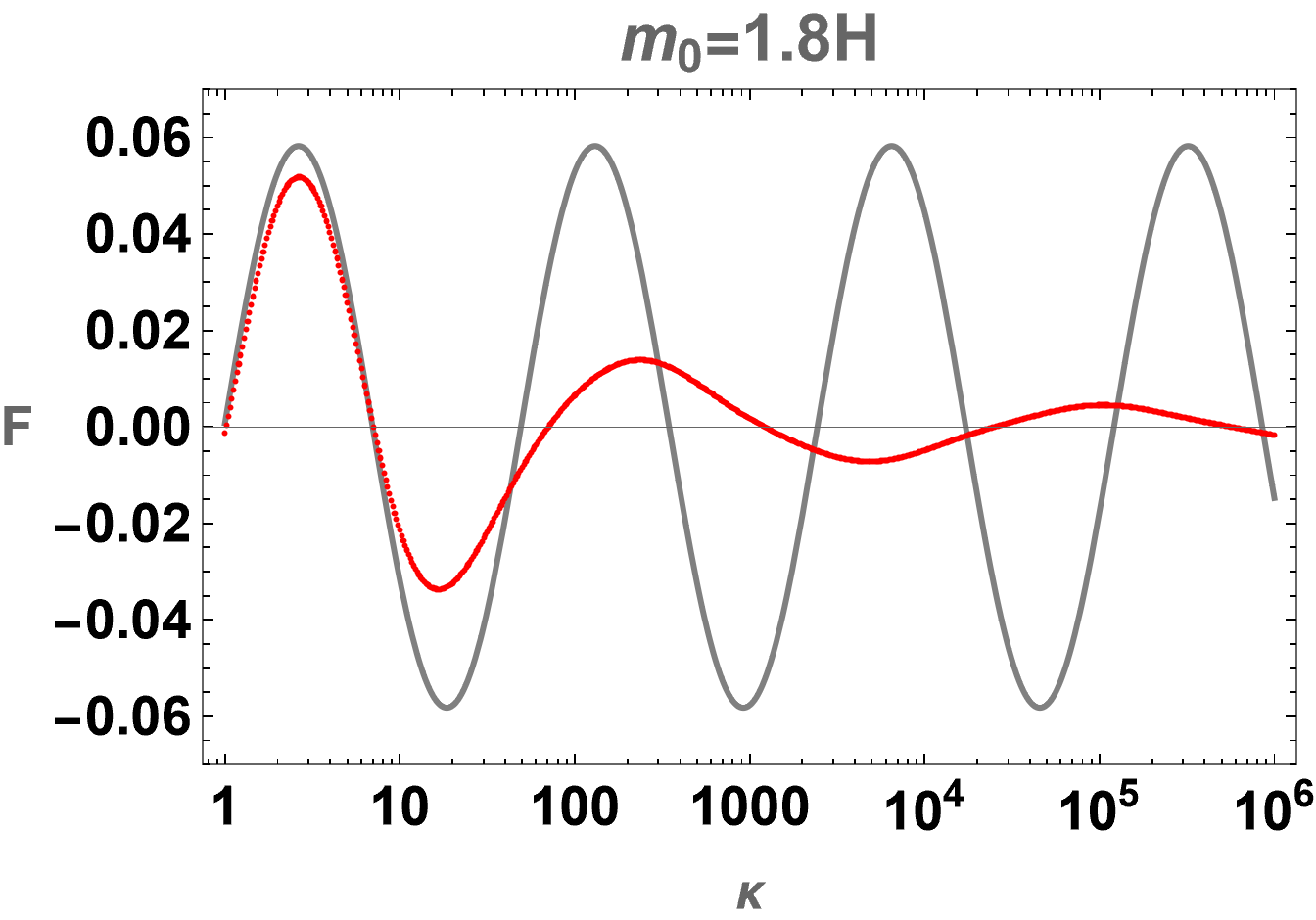}
  \end{center}
 \end{minipage}

  \begin{minipage}{0.5\hsize}
  \begin{center}
   \includegraphics[width=70mm]{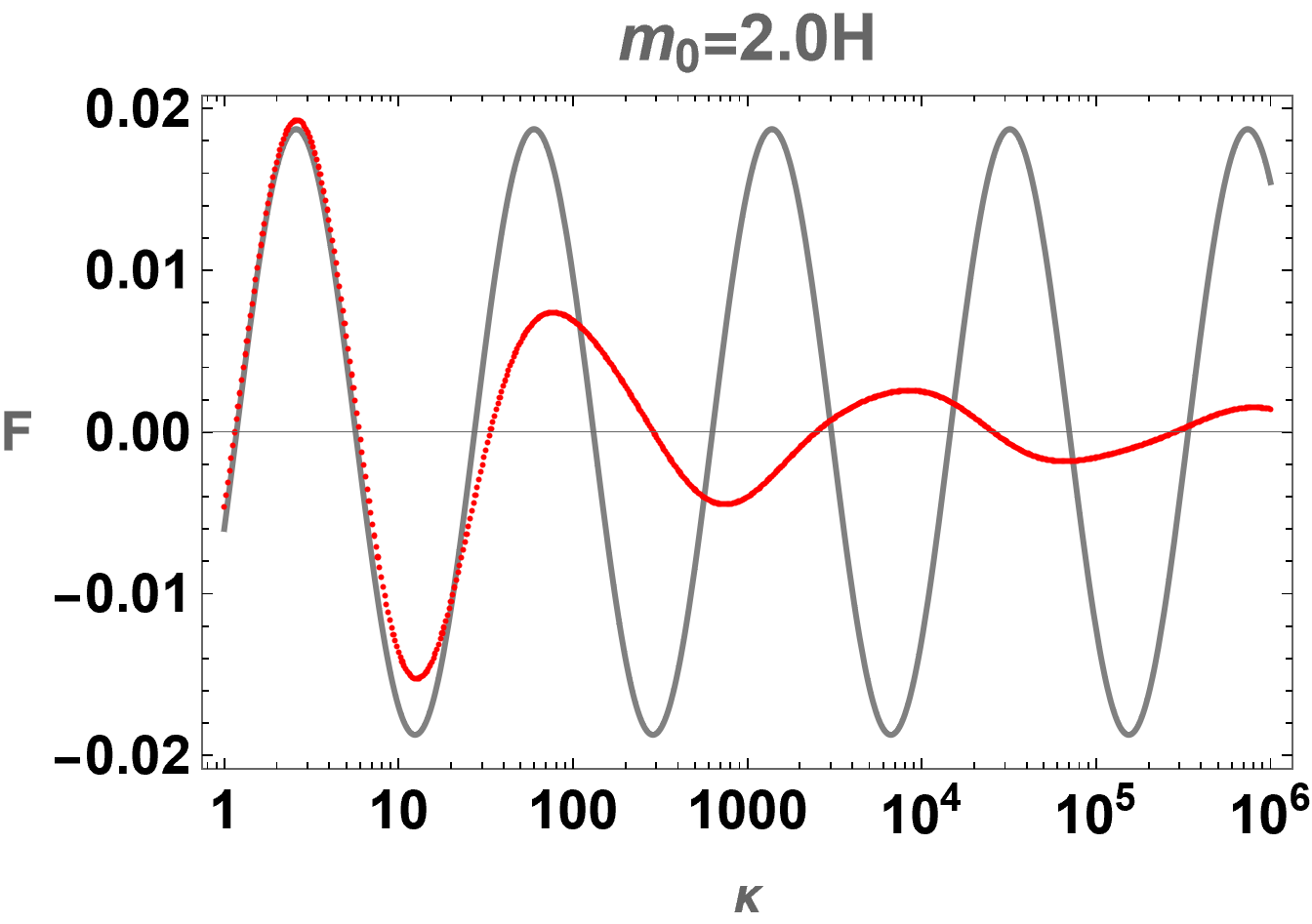}
  \end{center}
 \end{minipage}
 \begin{minipage}{0.5\hsize}
  \begin{center}
   \includegraphics[width=70mm]{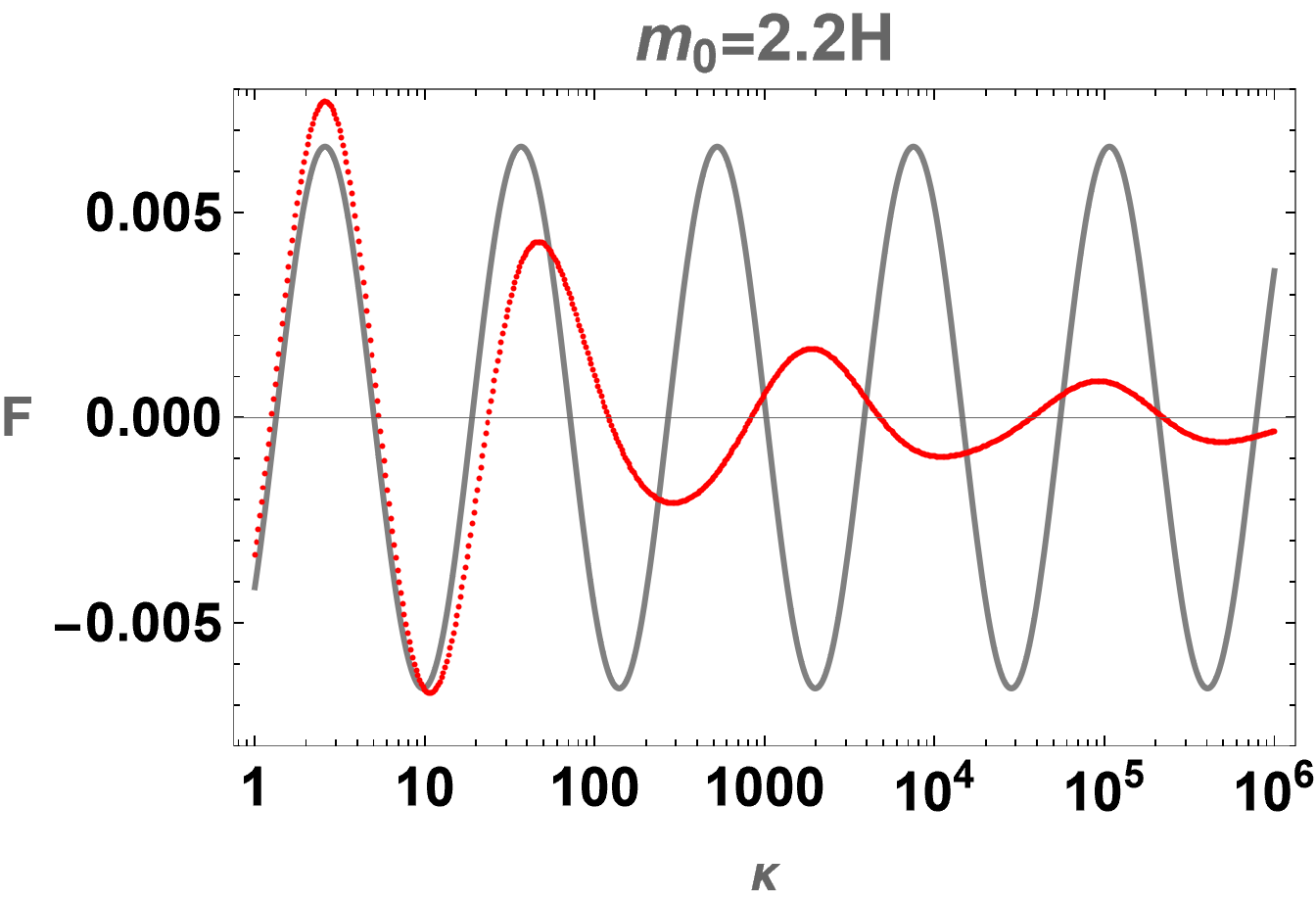}
  \end{center}
 \end{minipage}

\caption{Momentum  ($\kappa=k_1/k_3$) dependence of  $F_{\rm{conti.}}$ (red points). We set $m_0/H=1.6, 1.8, 2.0,$ and $2.2$. For comparison, we also show the particle case: $F_{\rm{particle}}$ with $m_{\sigma}/H=2.0, 2.2, 2.5,$ and $2.8$ (gray line).}
  \label{fig2}
\end{figure}

In Fig.~\ref{fig2}, we plot a momentum ($\kappa=k_1/k_3$) dependence of the function $F_{\rm{conti.}}$ for several values of $m_0$, $m_0/H=1.6, 1.8, 2.0,$ and $2.2$ (red points). There, for comparison, we also show the usual case with $F_{\rm{particle}}$ for $m_{\sigma}/H=2.0, 2.2, 2.5,$ and $2.8$, which gives a similar size as each $F_{\rm{conti}}$. For the mass integration in Eq.~\eqref{F}, we performed it numerically cutting the upper limit of the integral by some value, say $m^2/H^2=12$. This prescription does not affect the result significantly, and it is justified because contributions from heavy states are suppressed by the Boltzmann factor $e^{-\pi \mu}$.    
As can be seen from the figure, the result is striking: In the continuous case, the oscillations are damping as $\kappa$ becomes large (deep squeezed limit), while the amplitude is constant in the particle case. This kind of damping behavior is specific to the continuous case, and therefore, one can easily distinguish the continuous spectrum from the discrete one from the observation. Finally, we note that although we chose the gapped spectral density~\eqref{SD} for concreteness, this damping factor $1/\log \kappa$ comes from the mass integration of $\sin \left(m \log \kappa\right)$, so it should be irrelevant to the choice of $\rho(m^2)$.


\section{Summary}\label{Summary}
In this paper, we have investigated the effects of a massive field with a continuous spectrum on the bispectrum of the primordial curvature perturbations, or cosmological collider signal. The continuous spectrum is well motivated by several models with extra dimensions, and therefore, it is important to prepare templates of the non-Gaussianity for future observational searches.
Taking a simple model~\eqref{action}, we focused on the process where a single continuum isocurvaton is exchanged between the curvature perturbations as Fig.~\ref{fig1}. 
It turns out that the oscillating behavior of the bispectrum in the squeezed limit has a specific feature that the amplitude is damping in the deep squeezed limit. This is specific to the continuous case and cannot be mimicked by a discrete spectrum. Therefore, such damping oscillation signals can be evidence of the continuous spectrum on a very high energy scale.

Although here we focused on the simple process with a single continuum isocurvaton exchange for technical reasons, in principle, we can numerically evaluate contributions from more complicated diagrams with double or triple isocurvaton exchange, which can give a sizable signal to the bispectrum. The effects on the bispectrum at the loop level are also important for more wider application. Another interesting direction is to consider an explicit extra-dimensional setup for systems where an inflaton and a continuum isocurvaton coexist.
We leave all of them for future work.


\subsection*{Acknowledgements}
SA would like to thank Masahide Yamaguchi for helpful comments on the manuscript and Toshifumi Noumi, Lucas Pinol, Sebastien Renaux-Petel, and Fumiya Sano for collaboration on related works. 
This research was supported by the Chung-Ang University Research Grants in 2022, and the Basic Science Research Program
through the National Research Foundation (NRF) funded by the Ministry of Education,
Science and Technology in South Korea under the grant No. NRF-2022R1A2C2003567.

\begin{appendix}

\section{Perturbation in spatially flat gauge}\label{detail1}
Here we specify the action describing the perturbations around the inflationary background in the spatially flat gauge.

First, with Arnowitt-Deser-Misner (ADM) decomposition~\cite{Arnowitt:1962hi}, the four-dimensional metric can be decomposed as 
\begin{align}
g_{\mu\nu}=\left(\begin{array}{cc}
-N^2+N^iN_i & N_{i} \\
N_j & h_{ij}
\end{array}\right) ,\ \ 
g^{\mu\nu}=\left(\begin{array}{cc}
-\frac{1}{N^2} & \frac{N^{i}}{N^2} \\
\frac{N^{j}}{N^2} & h^{ij}-\frac{N^{i}N^j}{N^2}
\end{array}\right),  
\end{align}
where $N$, $N^i$, and $h_{ij}$ are lapse, shift vector, and induced metric on constant-$t$ hypersurface, respectively. The three-dimensional index $i$ is raised/lowered by $h_{ij}$ and its inverse, $h^{ij}$.

Then, in the spatially flat gauge (see Ref.~\cite{Wang:2013zva} for review), the fields and the metric are expanded as 
\begin{align}
&\phi=\phi_0(t)+\delta \phi, \ \ \sigma=\sigma_0+\delta\sigma, \label{expand1}\\
&N=1+\alpha, \ \ N_i=\partial_i\beta, \ \    h_{ij}=a^2(t)\delta_{ij}, \label{expand2}
\end{align}
where we omitted the vector- and tensor-perturbations.  

Now, we expand the action~\eqref{action} by Eqs.~\eqref{expand1} and~\eqref{expand2}. It turns out that $\alpha$ and $\beta$ appear as auxiliary fields without time derivatives, so that they can be integrated out~\cite{Maldacena:2002vr}. Their solutions to the first order in perturbations are\footnote{This is sufficient to discuss the expanded Lagrangian up to the third order in perturbations~\cite{Maldacena:2002vr}.} 
\begin{align}
&\alpha=\frac{1}{2 H \Mpl^2} f \dot{\phi}_0 \delta \phi,\\
&\partial^2\beta = -a^2 \frac{\dot{\phi}_0^2 f_{\sigma}}{2 H^2 \Mpl^2}\left[\frac{f}{f_{\sigma}} \frac{d}{d t}\left(\frac{H}{\dot{\phi}_0} \delta \phi\right)+H \delta \sigma\right], \label{beta}
\end{align}
where the derivative $\partial$ denotes the spatial one.
Inserting them back to the Lagrangian, we obtain the following second order Lagrangian:
\begin{align}
\nonumber \mathcal{L}_{2}=&\ \frac{a^3}{2}f(\delta \dot{\phi})^2-\frac{a}{2}  f(\partial \delta \phi)^2- a^3\left(\frac{V_{\phi\phi}}{2}-\left(3 \epsilon-\epsilon^2+\epsilon \eta\right) fH^2\right)  \delta \phi^2,\\
\nonumber &+\frac{a^3}{2}(\delta \dot{\sigma})^2-\frac{a}{2}  (\partial \delta \sigma)^2-\frac{a^3}{2}\left(U_{\sigma\sigma}-\frac{1}{2}f_{\sigma\sigma}\dot{\phi}_0^2\right)( \delta \sigma)^2\\
&+a^3 f_{\sigma} \dot{\phi}_0 \delta \sigma \delta \dot{\phi}-a^3 \epsilon f_{\sigma} \dot{\phi}_0 H  \delta \phi \delta \sigma.
\end{align}
In the same way, we get the third order Lagrangian: 
\begin{align}
\nonumber \mathcal{L}_{3}/a^3=&\ \frac{1}{2}f_{\sigma}\delta \sigma \left((\delta \dot{\phi})^2-\frac{1}{a^2}(\partial \delta \phi)^2\right)+\left(-\frac{1}{6} U_{\sigma \sigma \sigma}+\frac{1}{12}f_{\sigma\sigma\sigma}\dot{\phi}_0^2\right)(\delta \sigma)^3\\
\nonumber &+\left(-\frac{1}{6} V_{\phi \phi \phi}-\frac{1}{4 H \Mpl^2} f \dot{\phi}_0 V_{\phi \phi}+ \frac{3}{8H \Mpl^4} f^3 \dot{\phi}_0^3-\frac{1}{16 H^3 \Mpl^6} f^4 \dot{\phi}_0^5\right)(\delta \phi)^3\\
\nonumber &-\frac{f}{a^2} \delta \dot{\phi} \partial_i \beta \partial_i \delta \phi+\frac{1}{2} f_{\sigma \sigma}\dot{\phi}_0(\delta \sigma)^2  \delta \dot{\phi}+\frac{1}{2a^2 H \Mpl^2} f^2 \dot{\phi}_0^2(\delta \phi)^2 \partial^2 \beta\\
\nonumber &+\frac{1}{8 H^2 \Mpl^4} f_{\sigma}f^2 \dot{\phi}_0^4 \left(\delta \phi\right)^2 \delta \sigma+\frac{1}{4 H^2 \Mpl^4} f^3 \dot{\phi}^3_0(\delta \phi)^2 \delta \dot{\phi}- \frac{f_{\sigma}}{a^2} \dot{\phi}_0 \delta \sigma \partial_i \beta \partial_i \delta \phi\\
\nonumber &-\frac{1}{a^2} \delta \dot{\sigma} \partial_i \beta \partial_i \delta \sigma+\frac{1}{4a^4 H } f \dot{\phi}_0 \delta \phi\left(\partial^2 \beta\right)^2-\frac{1}{4a^4 H}f\dot{\phi}_0 \delta \phi\left(\partial_i \partial_j \beta\right)^2\\
\nonumber &-\frac{1}{4 a^2H \Mpl^2}  f^2 \dot{\phi}_0 \delta \phi(\partial \delta \phi)^2 -\frac{1}{4a^2 H \Mpl^2}  f \dot{\phi}_0 \delta \phi(\partial \delta \sigma)^2+\frac{1}{2a^2 H \Mpl^2} f^2 \dot{\phi}_0^2 \delta \phi \partial_i \beta \partial_i \delta \phi\\
\nonumber &-\frac{1}{4 H \Mpl^2}f\dot{\phi}_0\left( U_{\sigma \sigma}+\frac{1}{2}f_{\sigma \sigma} \dot{\phi}_0^2\right) \delta \phi(\delta \sigma)^2-\frac{1}{2 H \Mpl^2} f f_{\sigma} \dot{\phi}_0^2 \delta \phi \delta \dot{\phi} \delta \sigma\\
&-\frac{1}{4 H\Mpl^2} f^2 \dot{\phi}_0(\delta \dot{\phi})^2 \delta \phi -\frac{1}{4 H \Mpl^2}f\dot{\phi}_0 \delta \phi(\delta \dot{\sigma})^2,
\end{align}
where $\beta$ should be understood as Eq.~\eqref{beta} is inserted. Note that the results are consistent with Ref.~\cite{Chen:2009zp} by taking $f=({\rm{const.}}+\sigma)^2$.

\section{Derivation of bispectrum and squeezed limit}\label{detail2}
In this section, we show some details for deriving Eq.~(\ref{S}).

First, from the relevant interactions, \eqref{L_2z} and \eqref{L_3z}, we obtain the interaction Hamiltonian as follows,
\begin{align}
&H_{2, \rm{int}}(\tau) = \int d^3 \mathbf{x}\ 2a^3\epsilon\frac{f_{\sigma}}{f}\Mpl^2 H\delta \sigma \zeta^{\prime},\label{H_2}\\
&H_{3, \rm{int}}(\tau) = -\int d^3 \mathbf{x}\ a^2\epsilon \frac{f_{\sigma}}{f}\Mpl^2\delta \sigma \left(\zeta^{\prime 2}-(\partial \zeta)^2\right).\label{H_3}
\end{align}
We remind that the prime denotes the derivative with respect to the conformal time $\tau$.
Then, the bispectrum corresponding to the diagram in Fig.~\ref{fig1} can be calculated through the following formula~\cite{Wang:2013zva},
\begin{align}
\nonumber \langle\zeta_{\bf{k}_1} \zeta_{\bf{k}_2}\zeta_{\bf{k}_3}\rangle =  &\int_{-\infty}^{0}d\tau_1\int_{-\infty}^{0}d\tau_2 \langle 0|H^I_{\rm{int}}(\tau_1)\zeta^I_{\bf{k}_1} \zeta^I_{\bf{k}_2}\zeta^I_{\bf{k}_3}H^I_{\rm{int}}(\tau_2)| 0 \rangle \\
&-2{\rm{Re}}\int_{-\infty}^{0}d\tau_1\int_{-\infty}^{\tau_1}d\tau_2 \langle 0|\zeta^I_{\bf{k}_1} \zeta^I_{\bf{k}_2}\zeta^I_{\bf{k}_3}H^I_{\rm{int}}(\tau_1)H^I_{\rm{int}}(\tau_2)| 0 \rangle ,
\end{align}
where on the right-hand side we added a superscript $``I"$ on the fields and the interaction Hamiltonian to stress that they are described by the interaction picture. In both of the first and the second lines, one of $H^I_{\rm{int}}$ should be replaced by either of Eq.~\eqref{H_2} or Eq.~\eqref{H_3}, and the other $H^I_{\rm{int}}$ should be replaced by the rest. After some algebra, we obtain
\begin{align}
\nonumber &\langle\zeta_{\bf{k}_1} \zeta_{\bf{k}_2}\zeta_{\bf{k}_3}\rangle '\\
&=  -4 \epsilon^2 \Mpl^4H\left(\frac{f_{\sigma}}{f}\right)^2 \int \frac{d m^2}{2 \pi} \rho\left(m^2\right)\sum_{i=1}^{3}J^{(i)}_m(k_1,k_2,k_3)+{\rm{5\ permutations\ of\ }} {\bf{k}}_1,{\bf{k}}_2\ {\rm{and}}\ {\bf{k}}_3,\label{Bispectrum}
\end{align}
where 
\begin{align}
\nonumber J_{m}^{(1)}(k_1,k_2,k_3)=&\ {\rm{Re}} \biggl[\zeta_{k_1}^*\zeta_{k_2}^*\zeta_{k_3}(0)\int_{-\infty}^0d\tau_1a^2\left({\bf{k}_1}\cdot {\bf{k}_2}\zeta_{k_1}\zeta_{k_2}+\zeta_{k_1}^{\prime}\zeta_{k_2}^{\prime}\right)v_{k_3, m}(\tau_1)\\
&\times \int_{-\infty}^0d\tau_2a^3\zeta_{k_3}^{*\prime}v_{k_3,m}^{ *}(\tau_2)\biggr],\label{J1}\\
\nonumber J_{m}^{(2)}(k_1,k_2,k_3)=&-{\rm{Re}} \biggl[  \zeta_{k_1}\zeta_{k_2}\zeta_{k_3}(0) \int_{-\infty}^0d\tau_1a^2\left({\bf{k}_1}\cdot {\bf{k}_2}\zeta_{k_1}^*\zeta_{k_2}^*+\zeta_{k_1}^{*\prime}\zeta_{k_2}^{*\prime}\right)v_{k_3,m}(\tau_1)
\\
&\times \int_{-\infty}^{\tau_1}d\tau_2a^3\zeta_{k_3}^{*\prime}v_{k_3,m}(\tau_2)\biggr],\label{J2}\\
\nonumber J_{m}^{(3)}(k_1,k_2,k_3)=&-{\rm{Re}} \biggl[  \zeta_{k_1}\zeta_{k_2}\zeta_{k_3}(0) \int_{-\infty}^0d\tau_1a^3\zeta_{k_3}^{*\prime}v_{k_3,m}(\tau_1)
\\
&\times \int_{-\infty}^{\tau_1}d\tau_2a^2\left({\bf{k}_1}\cdot {\bf{k}_2}\zeta_{k_1}^*\zeta_{k_2}^*+\zeta_{k_1}^{*\prime}\zeta_{k_2}^{*\prime}\right)v_{k_3,m}(\tau_2)\biggr].\label{J3}
\end{align}
The prime on the left-hand side means that a momentum conservation factor $(2\pi)^3\delta^{(3)}({\bf{k}}_1+{\bf{k}}_2+{\bf{k}}_3)$ is extracted. $\zeta_k(\tau)$ and $v_{k,m}(\tau)$ are the mode functions for the curvature perturbation and the continuum isocurvaton respectively, 
\begin{align}
\zeta_k(\tau)=\frac{H}{\Mpl \sqrt{4 \epsilon k^3}}(1+i k \tau) e^{-i k \tau},
\end{align}
and
\begin{align}
v_{k,m}=-i e^{i\left(\nu+\frac{1}{2}\right) \frac{\pi}{2}} \frac{\sqrt{\pi}}{2} H(-\tau)^{3 / 2} H_{\nu}^{(1)}(-k \tau), \ \ {\rm{for}}\ \ m<3H/2,  
\end{align}
with $\nu \equiv \sqrt{\frac{9}{4}-\left(\frac{m}{H}\right)^2}$, and
\begin{align}
v_{k,m}=-i e^{-\frac{\pi}{2}\mu+i\frac{\pi}{4}} \frac{\sqrt{\pi}}{2} H(-\tau)^{3 / 2} H_{i\mu}^{(1)}(-k \tau), \ \ {\rm{for}}\ \ m>3H/2,  
\end{align}
with $\mu=\sqrt{\left(\frac{m}{H}\right)^2-\frac{9}{4}}$.

Now, let us discuss the squeezed limit ($k_1\sim k_2\gg k_3$) of Eq.~\eqref{Bispectrum}. To do so, it is useful to rewrite the time integrals in $J_m^{(i)}$ so that they are manifestly free from the IR divergence~\cite{Chen:2009zp}. Note that $J_m^{(i)}$ can be rewritten as 
\begin{align}
\sum_{i=1}^{3}J_m^{(i)}(k_1,k_2,k_3)+{\rm{5\ per.}}=\sum_{i=1}^{2}\tilde{J}_m^{(i)}(k_1,k_2,k_3)+{\rm{5\ per.}},
\end{align}
where 
\begin{align}
\nonumber \tilde{J}^{(1)}_m(k_1,k_2,k_3)= &-2\zeta_{k_1}^*\zeta_{k_2}^*\zeta_{k_3}(0)\int_{-\infty}^{0}d\tau_1\int_{-\infty}^{\tau_1}d\tau_2a^2(\tau_1)a^{3}(\tau_2) {\rm{Im}}\left[v_{k_3,m}(\tau_1)v_{k_{3},m}^{ *}(\tau_{2}) \zeta_{k_{3}}^{* \prime}(\tau_{2})\right]\\
&\times {\rm{Im}}\left[{\bf{k}_1}\cdot {\bf{k}_2}\zeta_{k_1}(\tau_1)\zeta_{k_2}(\tau_1)+\zeta_{k_1}'(\tau_1)\zeta_{k_2}'(\tau_1)\right] ,\label{J12}\\
\nonumber \tilde{J}^{(2)}_m(k_1,k_2,k_3)=& -2\zeta_{k_1}^*\zeta_{k_2}^*\zeta_{k_3}(0)\int_{-\infty}^{0}d\tau_1\int_{-\infty}^{\tau_1}d\tau_2a^{3}(\tau_1)a^2(\tau_2){\rm{Im}}\left[\zeta_{k_{3}}^{* \prime}(\tau_{1})\right]\\
&\times {\rm{Im}}\left[v_{k_{3},m}^{*}(\tau_{1})v_{k_3,m}(\tau_2)\left({\bf{k}_1}\cdot {\bf{k}_2}\zeta_{k_1}(\tau_2)\zeta_{k_2}(\tau_2)+\zeta_{k_1}'(\tau_2)\zeta_{k_2}'(\tau_2)\right)\right] .\label{J13}
\end{align}
Then, in the squeezed limit, 
we find that $\tilde{J}^{(1)}(k_1,k_2,k_3)=\tilde{J}^{(1)}(k_2,k_1,k_3)$ gives a dominant contribution, and it is explicitly given by 
\begin{align}
\nonumber &\tilde{J}_m^{(1)}(k_1,k_2,k_3) \\
\nonumber &\underset{k_1\sim k_2\gg k_3}{\simeq} \theta(m-3H/2) \frac{\pi H^3}{2^7\epsilon^3\Mpl^6} e^{-\pi\mu}k_1^{-9/2}k_3^{-3/2}{\rm{Im}}\left\{\left(\frac{k_3}{k_1}\right)^{-i \mu}J_{-}(i\mu)+\left(\frac{k_3}{k_1}\right)^{i \mu}J_{+}(i\mu)\right\}\\
&\ \ \ \ \ \ \ \ \ \ \ +\theta(3H/2-m) \frac{\pi H^3}{2^7\epsilon^3\Mpl^6}k_1^{-9/2+\nu}k_3^{-3/2-\nu}{\rm{Im}}(J_{-}(\nu)),
\end{align}
where $\theta(\cdot)$ is a unit step function, and
\begin{align}
\nonumber J_{\pm}(\nu)\equiv &\int_{-\infty}^{0} d x_{1}\int_{-\infty}^{0} d x_{2}\left(-x_{1}\right)^{-\frac{1}{2}\pm \nu}\left((1-2x_1^2)\sin 2 x_{1}-2 x_{1} \cos 2 x_{1}\right)\\
&\times  A^{\pm}(\nu)\left(-x_{2}\right)^{-\frac{1}{2}} H_{\nu}^{(1) *}\left(-x_{2}\right) e^{i x_{2}},\label{def_Jpm}
\end{align}
with 
\begin{align}
&A^-(\nu)=-i\frac{2^{\nu} \Gamma(\nu)}{\pi},\ \ A^+(\nu)=-i\frac{2^{-\nu} \Gamma(-\nu)}{\pi}e^{-i\pi\nu}. 
\end{align}
Therefore, we obtain
\begin{align}
\nonumber \langle\zeta_{\bf{k}_1} \zeta_{\bf{k}_2}\zeta_{\bf{k}_3}\rangle ' \underset{k_1\sim k_2\gg k_3}{\simeq} &-\frac{\pi H^4}{16 \epsilon \Mpl^2}\left(\frac{f_{\sigma}}{f}\right)^2 k_1^{-9/2}k_3^{-3/2} \\
\nonumber &\times \Biggl[ \int_{9H^2/4}^{\infty} \frac{d m^2}{2 \pi} \rho\left(m^2\right) e^{-\pi\mu}{\rm{Im}}\left\{\left(\frac{k_3}{k_1}\right)^{-i \mu}J_{-}(i\mu)+\left(\frac{k_3}{k_1}\right)^{i \mu}J_{+}(i\mu)\right\} \\
&+ \int_{0}^{9H^2/4} \frac{d m^2}{2 \pi} \rho\left(m^2\right)\left(\frac{k_3}{k_1}\right)^{-\nu}{\rm{Im}}(J_{-}(\nu))\Biggr].
\end{align}
Finally, with the definition of the shape function $S(k_1,k_2,k_3)$, 
\begin{align}
\langle\zeta_{\bf{k}_1} \zeta_{\bf{k}_2}\zeta_{\bf{k}_3}\rangle '=(2\pi)^4P_{\zeta}^2\frac{1}{(k_1k_2k_3)^2}S(k_1,k_2,k_3),    
\end{align}
where $P_{\zeta}=\frac{H^2}{8 \pi^2 \epsilon \Mpl^2}$ is the power spectrum of $\zeta$, 
we obtain 
\begin{align}
\nonumber S\underset{k_1\sim k_2\gg k_3}{\simeq} & -\frac{\pi}{4}\epsilon \Mpl^2 \left(\frac{f_{\sigma}}{f}\right)^2 \left(\frac{k_3}{k_1}\right)^{1/2} \\
\nonumber&\times \Biggl[ \int_{9H^2/4}^{\infty} \frac{d m^2}{2 \pi} \rho\left(m^2\right) e^{-\pi\mu}{\rm{Im}}\left\{\left(\frac{k_3}{k_1}\right)^{-i \mu}J_{-}(i\mu)+\left(\frac{k_3}{k_1}\right)^{i \mu}J_{+}(i\mu)\right\} \\
&+ \int_{0}^{9H^2/4} \frac{d m^2}{2 \pi} \rho\left(m^2\right)\left(\frac{k_3}{k_1}\right)^{-\nu}{\rm{Im}}(J_{-}(\nu))\Biggr]. \label{S_a}
\end{align}

Note that the integration in  $J_{\pm}$ in Eq.~\eqref{def_Jpm} can be performed exactly~\cite{Pinol:2021aun}, which results in 
\begin{align}
&J_+(i\mu)= -\frac{\sqrt{\pi} 2^{-1-2 i \mu}\left(e^{\pi \mu}-i\right) \Gamma\left(i \mu+\frac{7}{2}\right)(\operatorname{csch}(\pi \mu)+\operatorname{sech}(\pi \mu))}{(2 \mu-i) \Gamma(i \mu+1)}, \label{jp} \\
&J_-(i\mu)=J_+(-i\mu)e^{2\pi \mu},\label{jm}
\end{align}
for $m>3H/2$, and
\begin{align}
{\rm{Im}}(J_{-}(\nu))=-\frac{2^{-1+2 \nu} \Gamma\left(7 / 2-\nu\right) \Gamma\left(\nu\right) \cos \left(\pi \nu\right)}{\sqrt{\pi}\left(-1+2 \nu\right)\left(1+\sin \left(\pi \nu\right)\right)},
\end{align}
for $m<3H/2$.

For the specific choice of the spectral density~\eqref{SD} used in the main text, and for $m>m_0>3H/2$, the shape function~\eqref{S_a} with Eqs.~\eqref{jp} and \eqref{jm} reproduces the result~\eqref{S}.

\end{appendix}

\end{document}